\documentclass[12pt,english]{revtex4}
\usepackage[latin9]{inputenc}
\usepackage{color}
\usepackage{amstext}

\makeatletter

\providecommand{\tabularnewline}{\\}

\@ifundefined{textcolor}{}
{%
 \definecolor{BLACK}{gray}{0}
 \definecolor{WHITE}{gray}{1}
 \definecolor{RED}{rgb}{1,0,0}
 \definecolor{GREEN}{rgb}{0,1,0}
 \definecolor{BLUE}{rgb}{0,0,1}
 \definecolor{CYAN}{cmyk}{1,0,0,0}
 \definecolor{MAGENTA}{cmyk}{0,1,0,0}
 \definecolor{YELLOW}{cmyk}{0,0,1,0}
}

\usepackage{babel}
\def\apjl{ApJL }
\def\pasa{Proc. Astron. Soc. Aust.}
\def\nar{New Astronomy Reviews}
\def\mnras{MNRAS }

\def\apj{ApJ }

\def\aap{A\&A }
\def\nat{Nature }

\makeatother

\usepackage{babel}
\begin{document}

\title{Origin of the Galactic 511 keV emission from positrons produced inirregular supernovae}

\author{Hagai~B.~Perets}
\affiliation{Physics Department, Technion - Israel Insitute of Technology, Haifa 32000, Israel}

\begin{abstract}
Gamma ray emission of 511 keV lines arising from electron-positron
annihilation has been detected from the Galaxy since the 70s\cite{lev+78}.
Spatially resolved observations using the INTEGRAL satellite have
shown its full sky distribution to be strongly concentrated in the
Galactic bulge, with a smaller contribution from the disk, unlike
the situation in any other wavelength\cite{kno+05,bou+10,pra+11}.
The puzzling distribution of the positrons gave rise to various suggestions,
including stellar nucleosynthesis in core-collapse (CC) and type Ia
thermonuclear supernovae (SNe), accreting compact objects, and more
``exotic'' explanations of annihilation of dark-matter particles.
However, such models encounter difficulties in reproducing the total
Galactic 511 keV flux as well as its peculiar bulge-centered distribution\cite{pra+11}.
Theoretical models of SNe from thermonuclear Helium detonations on
white dwarfs (WDs) were suggested as potential additional sources
of positrons, contributing to the Galactic Gamma-ray emission\cite{woo+86}.
Here, we show that the recently discovered class of faint, calcium-rich
type Ib SNe (with the prototype SN 2005E)\cite{per+10}, thought to
arise from such explosions, also called ``.Ia'' SNs\cite{woo+86,bil+07},
can explain both the 511 keV flux and its distribution, and can eliminate
the need for non-astrophysical (e.g., dark matter annihilation) sources.
Such SNe comprise a fraction of only $\sim2\,\%$ of all SNe\cite{per+10,li+11b},
but they currently inject hundreds of times more positrons (from $^{44}{\rm Ti}$
decay) to the ISM than injected by CC SNe\cite{wal+11}, enough to
reproduce the total Galactic 511 keV line emission. They exclusively
explode in old environments\cite{per+10,per+11b,kas+12,lym+13,yua+13},
and their contribution to the 511 keV emission therefore follows the
old ($>10$ Gyrs) Galactic stellar population, dominated by the bulge,
thereby reproducing the observed large bulge-to-disk ratio. 
\end{abstract}
\maketitle
Best fit models of the observed full-sky distribution of 511 keV line
emission find a total Galactic flux of $17.1_{-1.5}^{+2.1}\times10^{-4}$
${\rm cm^{-2}s^{-1}}$ corresponding to a positron production rate
of $\dot{N}_{e^{+}}=19.6_{-2}^{+2.3}\times10^{42}$${\rm s}^{-1}$,
in a ``bulge+thick disk'' model\cite{wei+08b,bou+10} (following
the Galactic structure model in 
ref. \cite{pic+03}). The contribution from the extended bulge is
$\dot{N}_{e^{+}}^{bulge}=11.5_{-1.44}^{+1.8}\times10^{42}$${\rm s^{-1}}$
and the disk contribution amounts to $\dot{N}_{e^{+}}^{disk}=8.1_{-1.4}^{+1.5}\times10^{42}$${\rm s^{-1}}$,
giving a bulge-to-disk ratio $B/D$ of $\sim1.4$. The inferred injected
positron energy (or, equivalently, the mass of decaying and/or annihilating
dark matter particles in such models) from observations is constrained
to be to up to a few MeV\cite{pra+11}. Any consistent model for the
origin of the 511 keV line emission is required to reproduce the main
observations of Galactic emission, including the total positron production
rate, the bulge-to-disk production rate ratio, and the injected positron
energy. These constraints exclude or require fine-tuned/extreme conditions
for any of the suggested astrophysical and/or dark matter annihilation
origins of the Galactic 511 keV emission, and the origin of the $511$
keV $\gamma$-ray line has therefore remained a mystery (see 
ref. \cite{pra+11} for a detailed review).

Previous detailed radioactive models for positron production in SNe
have considered only CC SNe or regular type Ia thermonuclear explosions.
The former are expected to occur only in star forming regions, such
as the Galactic disk, while the latter are expected to occur in both
old and young stellar environments. However, studies of the delay
time distribution of type Ia SNe found it to be consistent with a
$t^{-1}$ behavior with the earliest SNe exploding only a few tens
of Myrs after the formation of their stellar progenitors; such SNe
are dominated by younger ($<1$ Gyr) stellar population\cite{mao+12}.
Therefore, these potential positron production sources cannot explain
the 511 keV distribution dominated by the old ($>10$ Gyrs) stellar
population of the bulge. Here, we show that accounting for the recently
discovered class of peculiar type Ib SNe, and their prototype SN 2005E,
can explain both the total positron production rate, as well as its
distribution.

SN-2005E like SNe\cite{per+10} are faint (typical $M_{b}\sim(-15)-(-16)$
absolute B magnitude), type Ib SNe, showing very strong calcium lines
and very little, if any, Iron at late nebular phases. The ejecta mass
of such SNe is inferred to be small, likely a few $0.1$ ${\rm M_{\odot}}$,
with very little mass of radioactive Nickel and Iron elements but
excessive amounts of Calcium (up to $\sim0.1$ ${\rm M_{\odot}}$
of Calcium inferred). Although type Ib SNe are typically observed
only in star-forming regions/galaxies, the majority of 2005E-like
SNe are observed in very old, possibly metal-poor, environments, such
as elliptical galaxies or galactic halos\cite{per+10,per+11b,kas+12,lym+13,yua+13},
with a typical stellar population older than 10 Gyrs. The SN properties
and environments suggest these are not CC SNe, but rather arise from
thermonuclear Helium detonations in/on WDs, which are not limited
to young environments\cite{per+10}. Based on observations and irrespective
of the specific model for these SNe, their distribution appears to
be most consistent with a very old stellar population\cite{per+10,per+11b,kas+12,lym+13,yua+13},
which is therefore adopted in the following. The inferred rate of
these SNe from volume-limited surveys is found to be $\sim10\pm5$
\% of the total rate of type Ia SNe \cite{per+10,li+11b}. These are
based on a small sample, hence the large uncertainty. However, these
rates are likely to be systematically and significantly underestimated,
due to detection bias against the discovery and follow-up observations
of SNe in galaxy bulges\cite{lea+11}. In the following we therefore
adopt $f_{05E}=10$ \% as the typical fractional rate, which serves
as a lower limit, to be better determined in future surveys. These
issues are discussed in more detail in the appendix. Current theoretical models of such explosions, sometimes termed
.Ia SNe\cite{bil+07}, can reproduce the general properties of these
SNe, such as their faint luminosities, early light curve (though the
faster rise of the modeled light curve compared to observations is
still not understood), and the excessive amount of intermediate elements
in their composition\cite{she+10,wal+11}. 

It has also been suggested that the large production of intermediate
elements in such SNe explains the excessive amount of Calcium observed
in Galaxy clusters\cite{per+10,mul+14}, providing further support
for these models. Overproduction of such intermediate elements is
typical to these explosions, due to nucleosynthetic freeze-out\cite{woo+86}.
In particular, Helium detonations on WDs lead to the large production
of radioactive Titanium isotopes\cite{woo+86}, $^{44}{\rm Ti}$,
with as much as $M_{^{44}{\rm Ti}}(.Ia)=3.3\times10^{-2}$ ${\rm M_{\odot}}$
of $^{44}{\rm Ti}$ produced in a single 2005E-like SN\cite{wal+11}
(based on theoretical models; the systematic uncertainties arising
from different models and their implications on the production rate
are discussed in the appendix). The eventual decay chain of these isotopes,
$^{44}{\rm Ti}\rightarrow^{44}{\rm Sc}\rightarrow^{44}{\rm Ca}$ ($\tau=85$
years), gives rise to the injection of positrons to the interstellar
medium (ISM), which would annihilate non-locally in the Galactic ISM
after diffusing in the ISM for typical timescales of $\sim10^{5}$
yrs\cite{pra+11}. Thereby, these can provide an additional positron
source for 511 keV line annihilation emission. As we show in the following,
accounting for these positron production channels (which potential
importance was first pointed out in Ref. \cite{woo+86}) from such
SNe in the Milky-Way galaxy shows that they could serve as the dominant
source of positrons producing the Galactic 511 keV line emission.

\begin{table}
\textbf{Production rate of $^{44}{\rm Ti}$ and frequencies of .Ia
and Core-collapse supernovae.}

\begin{tabular}{|c|c|c|c|}
\hline 
 & $M_{^{44}{\rm Ti}}$(${\rm M_{\odot})}$  & Fraction of Ia  & ${\rm R}_{{\rm SN}}$ (SNuM ; in Sbc galaxies)\tabularnewline
\hline 
\hline 
.Ia/2005E-like SNe  & $3.3\times10^{-2}$  & $f_{{\rm SN05E}}=0.1$  & $R_{{\rm SN05E}}=R_{{\rm Ia}}\times f_{{\rm SN05E}}=0.017\pm0.06$\tabularnewline
\hline 
Core-collapse  & $2\times10^{-4}$  & --  & ${\rm R_{CC}=}0.86$\tabularnewline
\hline 
\end{tabular}

\caption{Production rates of $^{44}{\rm Ti}$ production in .Ia SNe are based
on theoretical models in 
Ref. \cite{wal+11} (see discussion on their uncertainties in the
appendix), and production rates in CC SNe are based on observations of the
Cas-A and SN 1987A SNRs. The 2005E-like SNe fraction of Ia SNe is
based on 
refs. \cite{per+10} and \cite{li+11b}, and considered as a lower
limit (see appendix for further discussion); the overall SN rates are based
on 
ref. \cite{man+05}.}
\end{table}

\begin{table}
\textbf{Galaxy model of stellar components and star formation (SFR)
history}

\begin{tabular}{|c|c|c|c|c|}
\hline 
Galaxy Model  & Bulge (${\rm M_{\odot})}$  & Thin Disk (${\rm M_{\odot})}$  & Thick Disk (${\rm M_{\odot})}$  & Total Disk\tabularnewline
\hline 
\hline 
Mass  & $2.03\pm0.26\times10^{10}$  & $2.15\pm0.28\times10^{10}$  & $3.19\pm0.41\times10^{9}$  & $2.54\pm0.38\times10^{10}$\tabularnewline
\hline 
SFR history  & 10-11 Gyrs  & 0-11 Gyrs  & 10-11 Gyrs  & \tabularnewline
\hline 
\end{tabular}\caption{Galaxy structure is based on 
ref. \cite{rob+03},\textbf{\textcolor{blue}{{} }}which only provided
error bars for the bulge mass. The same level of uncertainty, $\sim13$
\%, is assumed for the disk components. SFR history is assumed to
be continuous in the given range; based on 
ref. \cite{wys+09}, and consistent with the Galaxy population SFR
modeling used in ref. \cite{rob+03}.}
\end{table}

The total positron production rate from 2005E-like SNe per $10^{10}$
Solar mass, $\dot{N}_{\text{e+,SN05E}}$, is 
\[
\dot{N}_{\text{e+,SN05E}}=n_{e+}(SN05E)\times R_{SN05E}=M_{^{44}{\rm Ti}}(.Ia)/(44\times M_{atom})\times R_{Ia}\times f_{05E},
\]
where $n_{e+}(SN05E)$ is the typical number of positrons injected
by a 2005E-like SN, ${\rm M_{{\rm atom}}}$ is the atomic mass, $R_{05E}=R_{Ia}\times f_{05E}$,
and $R_{Ia}$ are the rates of 2005E-like and type Ia SNe (SNuM, i.e.
the rate per $10^{10}$ ${\rm M_{\odot}}$ per century), respectively;
$f_{05E}$ is the fractional rate of 2005E-like SNe out of regular
type Ia SNe. For the calculations, we adopt the frequently used Galaxy
structure model in 
ref. \cite{rob+03}, which is also used for the best-fit modeling
of the 511 keV line distribution, i.e., the bulge and disks are similarly
defined. Note that in these models the bulge is defined differently
then in some of the typical denotations; it is defined to be the spherical
inner region the Galaxy, \textit{including} any older parts of the
stellar disk possibly extending to the inner regions. The disk is
then defined accordingly to have a hole in its center. These definitions
are adopted in all of the following. As prescribed in this model,
most of the stars in the Galaxy reside in the Galactic disk (total
mass of $2.54\times10^{10}$ ${\rm M_{\odot}}$, with $2.15\times10^{10}$${\rm M_{\odot}}$
in the thin disk and $3.19\times10^{9}$ ${\rm M_{\odot}}$ in the
thick disk (error bars were not given for these Galaxy component masses
in ref. \cite{rob+03} and we therefore adopt the same $\sim13$ \%
uncertainty level as estimated for the bulge mass determination in
the same ref.), with a smaller fraction in the bulge ($2.03\pm0.26\times10^{10}$),
and a much smaller, negligible fraction in the Galactic halo ($2.6\times10^{8}$
${\rm M_{\odot}}$). The typical age of the bulge, thick disk, and
halo stellar population is $>10$ Gyrs, and they are thought to have
formed over a period of $\sim$ 1 Gyr 
(see ref. \cite{wys+09} and references therein for an overview).
The thin disk stellar population is thought to have formed continuously
throughout the Galaxy evolution\cite{wys+09}. All of the bulge stellar
population (beside the negligible younger population in the Galactic
center) therefore contributes to the 2005E-like SN rate in the central
parts of the Galaxy. Outside the bulge, both the thick-disk population
and the oldest stellar population in the thin disk (where the latter
is$\sim10\%$ of the thin disk population, i.e., only stellar population
formed in the first Gyr, over the $\sim$10 Gyrs of evolution) contributes
to the 2005E-like SN rate. The old progenitors of the 2005E-like SNe
are therefore the main reason behind the dominant contribution of
the bulge to the 511 keV emission, rather than the more massive thin
disk. Since the 2005E-like SN rate is given in terms of the total
type Ia SN rate, we follow 
ref. \cite{pra+11} and adopt a rate of $R_{Ia}=0.17\pm0.06$ SNuM,
for $S_{bc}$ galaxies like the Milky-way\cite{man+05}.

We find $\dot{N}_{\text{e+,SN05E}}^{bulge}=9.74\pm3.66\times10^{42}{\rm s^{-1}},$$\dot{N}_{\text{e+,SN05E}}^{thin-disk}=1.05\pm0.39\times10^{42}{\rm s^{-1}},$$\dot{N}_{\text{e+,SN05E}}^{thick-disk}=1.95\pm0.73\times10^{42}{\rm s^{-1}},$
for a total of $\dot{N}_{\text{e+,SN05E}}^{MW}=\dot{N}_{\text{e+,SN05E}}^{bulge}+\dot{N}_{\text{e+,SN05E}}^{thin-disk}+\dot{N}_{\text{e+,SN05E}}^{thick-disk}=1.27\pm0.37\times10^{43}{\rm s^{-1}}$.
The observed emission of $^{26}{\rm Al}$ produced in CC SNe (and
seen only in the Galactic disk, not the bulge) also provides us with
an estimate of the expected contribution of positrons from such sources
for a total of $\dot{N}_{\text{e+,CC-SN}}^{disk}=3.5\pm0.26\times10^{42}$
\cite{wan+09,pra+11}. Taken together, we get $\dot{N}_{\text{e+}}^{MW}=12.7\pm3.7\times10^{42}{\rm s^{-1}}+3.5\pm0.26\times10^{42}=16.2\pm3.7\times10^{42}{\rm s^{-1}}$,
which is consistent with the total observed Galactic positron production.
The bulge-to-disk ratio is then $B/D=1.5\pm0.6$, also consistent
with the observed $B/D=1.42_{-0.30}^{+0.34}$. The relevant parameters
of the model and observations are summarized in Tables 1-3.

\begin{table}
\textbf{Positron production rate in the Galaxy: comparison of the
model with observations}

\begin{tabular}{|c|c|c|c|c|c|c|}
\hline 
 & \multicolumn{6}{c|}{$\dot{N}_{\text{e+,SN05E}}$$(10^{42}$${\rm s^{-1})}$}\tabularnewline
\hline 
 & bulge  & thin disk  & thick disk  & total disk  & total  & bulge/disk\tabularnewline
\hline 
\hline 
2005E-like/.Ia  & $9.74\pm3.66$  & $1.05\pm0.39$  & $1.95\pm0.73$  & $3.0\pm0.83$  & $12.7\pm3.7$  & \tabularnewline
\hline 
Contribution of $^{26}{\rm Al}$  & --  & $3.5\pm0.26$  & --  & $3.5\pm0.26$  & $3.5\pm0.26$  & \tabularnewline
\hline 
\noalign{\vskip\doublerulesep} \textbf{Model (total)}  & $9.74\pm3.66$  & $4.55\pm0.47$  & $1.95\pm0.73$  & $6.5\pm0.87$  & $16.2\pm3.7$  & $1.5\pm0.6$\tabularnewline
\hline 
\textbf{Observed $511$ keV}  & \textbf{$11.5_{-1.44}^{+1.8}$}  & \textbf{--}  & \textbf{--}  & \textbf{$8.1_{-1.4}^{+1.5}$}  & \textbf{$19.6_{-2}^{+2.3}$}  & \textbf{$1.42_{-0.30}^{+0.34}$}\tabularnewline
\hline 
\end{tabular}\caption{Comparison of the production rate inferred from observed 511 keV emission
and model prediction from 2005E-like SNe and and inferred contribution
from $^{26}{\rm Al}$ decay.}
\end{table}

The above calculation does not include positron production from $^{44}{\rm Ti}$
produced in CC SNe. $^{44}{\rm Ti}$ decay was observed in the supernovae
remnants (SNRs) Cas-A and 1987A (a total up to a few$\times10^{-4}$
${\rm M_{\odot}}$ in $^{44}{\rm Ti}$ per SN)\cite{iyu+94,gre+12}.
However, no gamma rays from $^{44}{\rm Ti}$ decay arising from the
expected (but optically obscured) SNe that should have gone off during
the past few 100 years in the galaxy have been detected. Detailed
models suggest these are atypical events\cite{ren+06,the+06,duf+13},
and therefore the total contribution from such sources is likely to
be small. Nevertheless, if these production rates are used as an upper
bound, i.e., taking the observed rates in Cas-A/1987A for \textit{all}
CC SNe, the positron production rate from such sources is $\sim3\times10^{42}{\rm s^{-1}}$\cite{pra+11},
which should be distributed mostly in the thin disk where current
star formation occurs (with $10\%$, $\sim0.3\times10^{42}$${\rm s^{-1}}$
contribution to the production rate from young stellar population
in the Galactic center). Together with the positron production from
the other sources discussed above, this can give rise to a total $\dot{N}_{\text{e+}}^{MW}=19\times10^{42}$${\rm s^{-1}}$
and $B/D=1.1$, still consistent with observations within the error-bars.
The uncertainty in the numbers given in Table 3 account for the known
uncertainties in the mass and distribution of the old and young stellar
population in the Galaxy, as well as the uncertainties in the type
Ia SNe rates. Additional discussions of other possible uncertainties
can be found in the appendix, most important is the
uncertainty in rate of 05E-like SNe, and its likely underestimate. 

Positron production from $^{44}{\rm Ti}$ decay in 2005E-like SNe
can provide a robust mechanism for the origin and distribution of
the Galactic 511 keV line emission in the Galaxy and can eliminate
the need for non-astrophysical (e.g., dark matter annihilation) or
fine-tuned astrophysical sources. This process also gives rise to
additional specific observable predictions. In particular, 2005E-like
SNe follow the oldest stellar population, and we therefore expect
the 511 keV emission to similarly be correlated with the old stellar
population in the Galaxy. In particular, beyond the expectation of
a large B/D, consistent with the observations, we expect a significant
fraction ($30\%$; see Table 2) of the disk 511 keV emission to originate
from the thick disk, rather than the more massive thin disk (i.e.,
a thick disk-to-thin disk positron production ratio of $0.3$), which
would not be expected from typical sources originating in the Galactic
center, such as the central massive black hole, or dark matter concentration.
In addition, given the observed delay-time distribution of standard
type Ia SNe ($\propto t^{-1}$), their expected current rate is only
$\sim5\%$ of the total type Ia rate, comparable with the current
rate of 2005E-like SNe ($\sim10\%$ of the total Ia rate). We therefore
expect a large fraction of any detected SNRs in the bulge (excluding
the younger Galactic center population in the $\sim100-200$ pcs central
environment) to be strong point sources of $^{44}{\rm Ti}$ decay
emission in $68/78$ keV emission lines, potentially detectable with
the current NuSTAR mission. These might even be observable in the
bulge of M31, and specifically in the SNR of SN 1885A, possibly related
to .Ia SNe\cite{per+11a} (although it is not a 2005E-like SN); such
observations are strongly encouraged.
\begin{acknowledgments}
We acknowledge support from the I-CORE Program of the Planning and
Budgeting Committee and The Israel Science Foundation (ISF) grant
1829/12, as well as the Technion Deloro fellowship and ISF BIKURA
fellowship.
\end{acknowledgments}
\bibliographystyle{naturemag}

\begin{thebibliography}{10}
\bibitem{lev+78} \bibinfo{author}{{Leventhal}, M.}, \bibinfo{author}{{MacCallum},
C.~J.} \& \bibinfo{author}{{Stang}, P.~D.} \newblock \bibinfo{title}{{Detection
of 511 keV positron annihilation radiation from the galactic center
direction}}. \newblock \emph{\bibinfo{journal}{\apjl}} \textbf{\bibinfo{volume}{225}},
\bibinfo{pages}{L11--L14} (\bibinfo{year}{1978}).

\bibitem{kno+05} \bibinfo{author}{{Kn{ö}dlseder}, J.} \emph{et~al.}
\newblock \bibinfo{title}{{The all-sky distribution of 511 keV
electron-positron annihilation emission}}. \newblock \emph{\bibinfo{journal}{A\&A}}
\textbf{\bibinfo{volume}{441}}, \bibinfo{pages}{513--532} (\bibinfo{year}{2005}).

\bibitem{bou+10} \bibinfo{author}{{Bouchet}, L.}, \bibinfo{author}{{Roques},
J.~P.} \& \bibinfo{author}{{Jourdain}, E.} \newblock \bibinfo{title}{{On
the morphology of the electron-positron annihilation emission as seen
by Spi/integral}}. \newblock \emph{\bibinfo{journal}{\apj}}
\textbf{\bibinfo{volume}{720}}, \bibinfo{pages}{1772--1780}
(\bibinfo{year}{2010}).

\bibitem{pra+11} \bibinfo{author}{{Prantzos}, N.} \emph{et~al.}
\newblock \bibinfo{title}{{The 511 keV emission from positron
annihilation in the Galaxy}}. \newblock \emph{\bibinfo{journal}{Reviews
of Modern Physics}} \textbf{\bibinfo{volume}{83}}, \bibinfo{pages}{1001--1056}
(\bibinfo{year}{2011}).

\bibitem{woo+86} \bibinfo{author}{{Woosley}, S.~E.}, \bibinfo{author}{{Taam},
R.~E.} \& \bibinfo{author}{{Weaver}, T.~A.} \newblock \bibinfo{title}{{Models
for type I supernova. I - Detonations in white dwarfs}}. \newblock
\emph{\bibinfo{journal}{ApJ}} \textbf{\bibinfo{volume}{301}},
\bibinfo{pages}{601--623} (\bibinfo{year}{1986}).

\bibitem{per+10} \bibinfo{author}{{Perets}, H.~B.} \emph{et~al.}
\newblock \bibinfo{title}{{A faint type of supernova from a white
dwarf with a helium-rich companion}}. \newblock \emph{\bibinfo{journal}{\nat}}
\textbf{\bibinfo{volume}{465}}, \bibinfo{pages}{322--325} (\bibinfo{year}{2010}).

\bibitem{bil+07} \bibinfo{author}{{Bildsten}, L.} \emph{et~al.}
\newblock \bibinfo{title}{{Faint thermonuclear supernovae from
AM Canum Venaticorum binaries}}. \newblock \emph{\bibinfo{journal}{ApJL}}
\textbf{\bibinfo{volume}{662}}, \bibinfo{pages}{L95--L98} (\bibinfo{year}{2007}).

\bibitem{li+11b} \bibinfo{author}{{Li}, W.} \emph{et~al.} \newblock
\bibinfo{title}{{Nearby supernova rates from the Lick Observatory
Supernova Search - II. The observed luminosity functions and fractions
of supernovae in a complete sample}}. \newblock \emph{\bibinfo{journal}{\mnras}}
\textbf{\bibinfo{volume}{412}}, \bibinfo{pages}{1441--1472}
(\bibinfo{year}{2011}).

\bibitem{wal+11} \bibinfo{author}{{Waldman}, R.} \emph{et~al.}
\newblock \bibinfo{title}{{Helium Shell Detonations on low-mass
white dwarfs as a possible explanation for SN 2005E}}. \newblock
\emph{\bibinfo{journal}{\apj}} \textbf{\bibinfo{volume}{738}},
\bibinfo{pages}{21} (\bibinfo{year}{2011}).

\bibitem{per+11b} \bibinfo{author}{{Perets}, H.~B.} \emph{et~al.}
\newblock \bibinfo{title}{{The old environment of the faint Calcium-rich
supernova SN 2005cz}}. \newblock \emph{\bibinfo{journal}{\apjl}}
\textbf{\bibinfo{volume}{728}}, \bibinfo{pages}{L36} (\bibinfo{year}{2011}).

\bibitem{kas+12} \bibinfo{author}{{Kasliwal}, M.~M.} \emph{et~al.}
\newblock \bibinfo{title}{{Calcium-rich gap transients in the
remote outskirts of Galaxies}}. \newblock \emph{\bibinfo{journal}{\apj}}
\textbf{\bibinfo{volume}{755}}, \bibinfo{pages}{161} (\bibinfo{year}{2012}).

\bibitem{lym+13} \bibinfo{author}{{Lyman}, J.~D.} \emph{et~al.}
\newblock \bibinfo{title}{{Environment-derived constraints on
the progenitors of low-luminosity Type I supernovae}}. \newblock
\emph{\bibinfo{journal}{\mnras}} \textbf{\bibinfo{volume}{434}},
\bibinfo{pages}{527--541} (\bibinfo{year}{2013}).

\bibitem{yua+13} \bibinfo{author}{{Yuan}, F.} \emph{et~al.}
\newblock \bibinfo{title}{{Locations of peculiar supernovae as
a diagnostic of their origins}}. \newblock \emph{\bibinfo{journal}{\mnras}}
\textbf{\bibinfo{volume}{432}}, \bibinfo{pages}{1680--1686}
(\bibinfo{year}{2013}).

\bibitem{wei+08b} \bibinfo{author}{{Weidenspointner}, G.} \emph{et~al.}
\newblock \bibinfo{title}{{Positron astronomy with SPI/INTEGRAL}}.
\newblock \emph{\bibinfo{journal}{\nar}} \textbf{\bibinfo{volume}{52}},
\bibinfo{pages}{454--456} (\bibinfo{year}{2008}).

\bibitem{pic+03} \bibinfo{author}{{Robin}, A.~C.}, \bibinfo{author}{{Reyl{é}},
C.}, \bibinfo{author}{{Derri{è}re}, S.} \& \bibinfo{author}{{Picaud},
S.} \newblock \bibinfo{title}{{A synthetic view on structure
and evolution of the Milky Way}}. \newblock \emph{\bibinfo{journal}{\aap}}
\textbf{\bibinfo{volume}{409}}, \bibinfo{pages}{523--540} (\bibinfo{year}{2003}).

\bibitem{mao+12} \bibinfo{author}{{Maoz}, D.} \& \bibinfo{author}{{Mannucci},
F.} \newblock \bibinfo{title}{{Type-Ia supernova rates and the
progenitor problem: A review}}. \newblock \emph{\bibinfo{journal}{\pasa}}
\textbf{\bibinfo{volume}{29}}, \bibinfo{pages}{447--465} (\bibinfo{year}{2012}).

\bibitem{lea+11}\bibinfo{author}{{Leaman}, J.} \& \bibinfo{author}{{Li},
W.} \& \bibinfo{author}{{Chornock}, R.} \& \bibinfo{author}{{Filippenko},
A. V.} \newblock \bibinfo{title}{{Nearby supernova rates from
the Lick Observatory Supernova Search - I. The methods and data base}}.
\newblock \emph{\bibinfo{journal}{\mnras}} \textbf{\bibinfo{volume}{412}},
\bibinfo{pages}{1419--1440} (\bibinfo{year}{2011}).

\bibitem{she+10} \bibinfo{author}{{Shen}, K.~J.} \emph{et~al.}
\newblock \bibinfo{title}{{Thermonuclear .Ia supernovae from helium
shell detonations: Explosion models and observables}}. \newblock
\emph{\bibinfo{journal}{\apj}} \textbf{\bibinfo{volume}{715}},
\bibinfo{pages}{767--774} (\bibinfo{year}{2010}).

\bibitem{mul+14} \bibinfo{author}{{Mulchaey}, J.~S.}, \bibinfo{author}{{Kasliwal},
M.~M.} \& \bibinfo{author}{{Kollmeier}, J.~A.} \newblock
\bibinfo{title}{{Calcium-rich gap transients: Solving the Calcium
conundrum in the intracluster medium}}. \newblock \emph{\bibinfo{journal}{\apjl}}
\textbf{\bibinfo{volume}{780}}, \bibinfo{pages}{L34} (\bibinfo{year}{2014}).

\bibitem{rob+03} \bibinfo{author}{{Robin}, A.~C.}, \bibinfo{author}{{Reyl{é}},
C.}, \bibinfo{author}{{Derri{è}re}, S.} \& \bibinfo{author}{{Picaud},
S.} \newblock \bibinfo{title}{{A synthetic view on structure
and evolution of the Milky Way}}. \newblock \emph{\bibinfo{journal}{\aap}}
\textbf{\bibinfo{volume}{409}}, \bibinfo{pages}{523--540} (\bibinfo{year}{2003}).

\bibitem{wys+09} \bibinfo{author}{{Wyse}, R.~F.~G.} \newblock
\bibinfo{title}{{The star-formation history of the Milky Way Galaxy}}.
\newblock In \bibinfo{editor}{{Mamajek}, E.~E.}, \bibinfo{editor}{{Soderblom},
D.~R.} \& \bibinfo{editor}{{Wyse}, R.~F.~G.} (eds.) \emph{\bibinfo{booktitle}{IAU
Symposium}}, vol. \bibinfo{volume}{258} of \emph{\bibinfo{series}{IAU
Symposium}}, \bibinfo{pages}{11--22} (\bibinfo{year}{2009}).

\bibitem{man+05} \bibinfo{author}{{Mannucci}, F.} \emph{et~al.}
\newblock \bibinfo{title}{{The supernova rate per unit mass}}.
\newblock \emph{\bibinfo{journal}{\aap}} \textbf{\bibinfo{volume}{433}},
\bibinfo{pages}{807--814} (\bibinfo{year}{2005}).

\bibitem{wan+09} \bibinfo{author}{{Wang}, W.} \emph{et~al.}
\newblock \bibinfo{title}{{Spectral and intensity variations of
Galactic $^{26}$Al emission}}. \newblock \emph{\bibinfo{journal}{\aap}}
\textbf{\bibinfo{volume}{496}}, \bibinfo{pages}{713--724} (\bibinfo{year}{2009}).

\bibitem{iyu+94} \bibinfo{author}{{Iyudin}, A.~F.} \emph{et~al.}
\newblock \bibinfo{title}{{COMPTEL observations of Ti-44 gamma-ray
line emission from CAS A}}. \newblock \emph{\bibinfo{journal}{\aap}}
\textbf{\bibinfo{volume}{284}}, \bibinfo{pages}{L1--L4} (\bibinfo{year}{1994}).

\bibitem{gre+12} \bibinfo{author}{{Grebenev}, S.~A.}, \bibinfo{author}{{Lutovinov},
A.~A.}, \bibinfo{author}{{Tsygankov}, S.~S.} \& \bibinfo{author}{{Winkler},
C.} \newblock \bibinfo{title}{{Hard-X-ray emission lines from
the decay of $^{44}$Ti in the remnant of supernova 1987A}}. \newblock
\emph{\bibinfo{journal}{\nat}} \textbf{\bibinfo{volume}{490}},
\bibinfo{pages}{373--375} (\bibinfo{year}{2012}). \newblock
\eprint{1211.2656}.

\bibitem{ren+06} \bibinfo{author}{{Renaud}, M.} \emph{et~al.}
\newblock \bibinfo{title}{{An INTEGRAL/IBIS view of young Galactic
SNRs through the $^{44}$Ti gamma-ray lines}}. \newblock \emph{\bibinfo{journal}{\nar}}
\textbf{\bibinfo{volume}{50}}, \bibinfo{pages}{540--543} (\bibinfo{year}{2006}).

\bibitem{the+06} \bibinfo{author}{{The}, L.-S.} \emph{et~al.}
\newblock \bibinfo{title}{{Are $^{44}$Ti-producing supernovae
exceptional?}} \newblock \emph{\bibinfo{journal}{A\&A}} \textbf{\bibinfo{volume}{450}},
\bibinfo{pages}{1037--1050} (\bibinfo{year}{2006}).

\bibitem{duf+13} \bibinfo{author}{{Dufour}, F.} \& \bibinfo{author}{{Kaspi},
V.~M.} \newblock \bibinfo{title}{{Limits on the number of galactic
young supernova remnants emitting in the decay lines of $^{44}$Ti}}.
\newblock \emph{\bibinfo{journal}{\apj}} \textbf{\bibinfo{volume}{775}},
\bibinfo{pages}{52} (\bibinfo{year}{2013}). \newblock \eprint{1308.4859}.

\bibitem{per+11a} \bibinfo{author}{{Perets}, H.~B.}, \bibinfo{author}{{Badenes},
C.}, \bibinfo{author}{{Arcavi}, I.}, \bibinfo{author}{{Simon},
J.~D.} \& \bibinfo{author}{{Gal-yam}, A.} \newblock \bibinfo{title}{{An
emerging class of bright, fast-evolving supernovae with low-mass ejecta}}.
\newblock \emph{\bibinfo{journal}{\apj}} \textbf{\bibinfo{volume}{730}},
\bibinfo{pages}{89} (\bibinfo{year}{2011}).

\bibitem{pic+04} \bibinfo{author}{{Picaud}, S.} \& \bibinfo{author}{{Robin},
A.~C.} \newblock \bibinfo{title}{{3D outer bulge structure from
near infrared star counts}}. \newblock \emph{\bibinfo{journal}{\aap}}
\textbf{\bibinfo{volume}{428}}, \bibinfo{pages}{891--903} (\bibinfo{year}{2004}).

\bibitem{mcm11} \bibinfo{author}{{McMillan}, P.~G.} \newblock
\bibinfo{title}{{Mass models of the Milky Way}}. \newblock \emph{\bibinfo{journal}{\mnras}}
\textbf{\bibinfo{volume}{414}}, \bibinfo{pages}{2446--2457}
(\bibinfo{year}{2011}).

\bibitem{dso+14} \bibinfo{author}{{D'Souza}, R.}, \bibinfo{author}{{Kauffmann},
G.} \bibinfo{author}{{Wang}, J.} \& \bibinfo{author}{{Vegetti},
S.} \newblock \bibinfo{title}{{Parametrizing the Stellar Haloes
of Galaxies\textbf{}}, \newblock \emph{\bibinfo{journal}{arXiv}}
\textbf{\bibinfo{volume}{arxiv:1404.2123}}, (\bibinfo{year}{2014}).

\bibitem{li+11} \bibinfo{author}{{Li}, W.} \emph{et~al.} \newblock
\bibinfo{title}{{Nearby supernova rates from the Lick Observatory
Supernova Search - III. The rate-size relation, and the rates as a
function of galaxy Hubble type and colour}}. \newblock \emph{\bibinfo{journal}{\mnras}}}.
\textbf{\bibinfo{volume}{412}}, \bibinfo{pages}{1473--1507}
(\bibinfo{year}{2011}).

\bibitem{she+13} \bibinfo{author}{{Shen}, K.~J.} \& \bibinfo{author}{{Bildsten},
L.} \newblock \bibinfo{title}{{The ignition of carbon detonations
via converging shock waves in White Dwarfs}}. \newblock \emph{\bibinfo{journal}{ArXiv:1305.6925}}
\textbf{\bibinfo{volume}{1}} (\bibinfo{year}{2013}).

\bibitem{mil+02} \bibinfo{author}{{Milne}, P.~A.}, \bibinfo{author}{{Kurfess},
J.~D.}, \bibinfo{author}{{Kinzer}, R.~L.} \& \bibinfo{author}{{Leising},
M.~D.} \newblock \bibinfo{title}{{Supernovae and positron annihilation
radiation}}. \newblock \emph{\bibinfo{journal}{\nar}} \textbf{\bibinfo{volume}{46}},
\bibinfo{pages}{553--558} (\bibinfo{year}{2002}).

\end{thebibliography}

\section*{Appendix }

The calculated rates for Galactic positron production depend on various
SN characteristics and rates, as well as the Galaxy structure. Here
we discuss these uncertainties in more detail.

\subsection{Bulge mass}

The best fit modeling of the observed distribution of the 511 keV
line emission\cite{pra+11} made, makes use of the Galactic structure
constructed in ref. \cite{rob+03}. Later modeling of the Galaxy structure\cite{pic+04}
suggest a bulge mass of $2.4\pm0.6\times10^{10}$ ${\rm M_{\odot}}$,
which is also consistent with more recent models\cite{mcm11}. These
latter models are consistent with the most frequently used model for
the Galaxy in ref. \cite{rob+03}, which was adopted throughout this
paper. Note that the larger estimates allow the bulge mass to be as
much as $1.5$ larger than the adopted model (in which ${\rm M_{{\rm bulge}}=2\times10^{10}}$
${\rm M_{\odot}})$, and would therefore correspond to a larger positron
production from 2005E-like SNe (still consistent with observations),
and possibly leading to an even larger B/D ratio, even if the disk
mass is similarly up-scaled (since the total contribution from $^{26}{\rm Al}$
is known from observations independently of the disk mass, it would
not be similarly up-scaled, and would therefore contribute a smaller
fraction of the disk positrons in this case).

\subsection{SN rate estimates and distribution}

SN rate are notoriously difficult to measure, since they require an
accurate and detailed knowledge of surveys used and their control
time\cite{mao+12}. As can be inferred from our calculations in the
main text, uncertainties in SN rate estimates dominate the overall
uncertainties in the positron production rates we derive. Here we
adopted the same type Ia SN rate as in Ref. \cite{pra+11}, following
ref. \cite{man+05}, for type Ia SN. These are generally consistent
with the recent rate estimates based on a larger local SN sample in
ref. \cite{li+11}. The formal inferred rate of 2005E-like SNe is
$10\pm5\%$ of the type Ia SN; it is based on the same \cite{li+11}
SN sample\cite{li+11b}. This rate is also consistent with the suggestions
that 2005E-like SNe produce most of the Calcium in the intra-cluster
medium\cite{per+10}, in which case the best estimate for their rate
can be as high as $16\%$ of the type Ia SN rate\cite{mul+14}.\textcolor{blue}{{}
}However, most of the old stellar populations in galaxies (such as
the stellar populations of elliptical galaxies and stellar halos in
which such SNe were observed) is expected to reside in galaxy bulges,
thick disks, galactic halos (external to the bulge) and the intergalactic
medium in galaxy clusters, depending on the galaxy type. 

\begin{table}
\textbf{Host and spatial location of 2005E-like SNe}

\begin{tabular}{|c|c|c|c|c|}
\hline 
\textbf{SN} & \textbf{Closest galaxy} & \textbf{Galaxy type} & \textbf{Location$^{1}$} & \textbf{Offset$^{2}$ }\tabularnewline
\hline 
PTF09dav & Anon & disturbed Sb  & IGM & 40\tabularnewline
\hline 
2010et/PTF 10iuv & CGCG 170-011 & E & IGM & 37\tabularnewline
\hline 
PTF11bij & IC 3956  & E & halo/IGM & 33\tabularnewline
\hline 
2000ds & NGC 2768 & E/S0 & halo/X-bulge & \tabularnewline
\hline 
2005cz & NGC 4589 & E & halo/X-bulge & \tabularnewline
\hline 
2007ke & NGC 1129 & E & halo/X-bulge & \tabularnewline
\hline 
2012hn & NGC 2272 & E/S0 & halo/X-bulge & \tabularnewline
\hline 
2005E & NGC 1032 & S0/a & halo/X-bulge & \tabularnewline
\hline 
2001co & NGC 5559 & SBb & disk & \tabularnewline
\hline 
2003H & NGC 2207+IC 2163 & Interacting spirals & interacting disks & \tabularnewline
\hline 
2003dg & UGC 6934 & Scd & disk & \tabularnewline
\hline 
2003dr & NGC 5714 & Scd & disk & \tabularnewline
\hline 
\end{tabular}

\begin{raggedright}
$^{1}$Offset from center of nearest galaxy (kpc) for IGM SNe
\par\end{raggedright}

\raggedright{}$^{2}$Location: central bulge ($<$3 kpc from center),
disk ($<$1 kpc from disk), halo/X-bulge, IGM
\end{table}

The detection efficiency in the inner bulges of galaxies is very poor,
and significantly limits or even completely prohibits the discovery
of SNe in these regions\cite{lea+11}. Galactic halos contain a few
up-to 25\% of the stellar mass of galaxies in late type galaxies (Sb
and later, such as the Milky-way), and between 30-70\% of the stellar
mass in early type galaxies (E/S0/Sa)\cite{dso+14}. Together, and
given the bias against detection of SNe in galaxy bulges/inner regions,
one would expect to locate SNe from old stellar populations in either
the thick disk of late type galaxies, the halo of early type galaxies,
or in the intergalactic medium of galaxy clusters. Table 4 show that
the distribution of 2005E-like SNe locations in galaxies, as well
as the type of galaxies are consistent with these expectations, i.e.
they originate in very old stellar populations. Given the thick-disk
to bulge ratio in late type galaxies, as well as the relative stellar
mass in halos, compared with the stellar mass in the inner regions
of early type galaxies, we expect between 1/2 to 2/3 of 2005E-like
SNe to be missed due to the bias against bulge SNe. The rate inferred
from the detected SNe is therefore highly likely to be systematically
underestimated by a factor of 2-3 and the formal uncertainty in their
rates can not be taken at face value. Instead, we adopt the overall
rate of $10\%$ of the type Ia SNe, which is the formally inferred
from the \textit{detected} 2005E-like SNe, but caution that it should
be considered as a lower limit. This is also consistent with the study
in ref. \cite{yua+13} showing that the distribution of these SNe
is consistent with an old ($>10$ Gyr), likely metal poor, stellar
population as typically observed in galaxy bulges and stellar halos,
once (simplistically) accounting for the detection bias against observations
of the SNe in the inner regions of galaxies.

\subsection{$^{44}$Ti production rate in thermonuclear Helium-detonations on
WDs}

The production rate of $^{44}{\rm Ti}$ in 05E-like SNe can not be
directly constrained by current observations; though, as suggested
in the main text, SNRs of such SNe, once discovered, should show strong
emission from $^{44}{\rm Ti}$ decay, which would enable an observational
calibration of the positron yield from these SNe. 

As discussed in detail in ref. \cite{per+10}, the various properties
of 05E-like SNe (old environment, excessive production of Calcium,
low ejecta-masses and low luminosity/low production of $^{56}{\rm Ni}$),
all point out to an origin from a Helium-detonation on a WD. In such
explosions explosive-nucleosynthesis occurs through the alpha-chain
process, in which heavier elements are produced through consecutive
fusion of alpha nuclei with progressively heavier atoms. In particular,
$^{44}{\rm Ti}$ is directly produced from the fusion of an alpha
nuclei with $^{40}{\rm Ca}$ atoms (i.e. they are neighboring isotopes
in the alpha-chain), and therefore it would require a highly fine-tuned
conditions in order to produce large abundances of one of these isotopes
without producing significant abundances of the other. Indeed, the
fractional ratio of $^{44}{\rm Ti}$ to stable Calcium, $^{40}{\rm Ca}$
is found to be large in all studied cases of Helium-detonations. Calcium
is known observationally to be produced excessively (hence these SNe
are sometimes called Calcium-rich SNe), with an estimated mass of
$\sim0.1$ ${\rm M_{\odot}}$ of Calcium in the case of SN 2005E\cite{per+10}.
By taking the range of $^{44}{\rm Ti}$ to $^{40}{\rm Ca}$ ratios
found in theoretical models and the estimated Calcium mass in SN-2005E
one can infer that large $^{44}{\rm Ti}$ abundances should be produced. 

Though the exact model for 05E-like SNe is still debated, the above
discussion suggests a robust production of $^{44}{\rm Ti}$ in these
SNe, and we therefore make use of the theoretical models for Helium-detonations
in refs. \cite{per+10,she+10,wal+11} as proxies to infer the rate
of $^{44}{\rm Ti}$ production. In \cite{per+10} we ran a grid of
1-zone helium detonations, with various conditions for temperatures,
densities and masses, independent of a specific spatial model (e.g.
Helium-shell detonation, Helium WD merger with CO-WD etc.). The models
in ref. \cite{she+10,wal+11} include a detonation of a helium shell
of varying mass between 0.15-0.3 ${\rm M_{\odot}}$ on a CO-WD with
masses ranging between 0.45- 1.2 ${\rm M_{\odot}}$. All the models
predict the production of very large abundance of $^{44}{\rm Ti}$
typically in the range of $5\times10^{-3}-3.3\times10^{-2}$${\rm M_{\odot}}$.
Later studies suggest that more massive CO WDs are likely to detonate
and explode due to the Helium-shell ignition\cite{she+13}, and are
not likely to produce 2005E-like SNe, where as lower mass WD may only
have the helium-shell detonation, thereby producing .Ia events. Of
the latter, the more likely models for 2005E-like SNe are therefore
those with the lowest mass CO WDs; most of which produce $>3\times10^{-2}$${\rm M_{\odot}}$
of $^{44}{\rm Ti}$. In particular, the best-fit model compared with
the observations of SN 2005E light curve is the model with $0.45$
${\rm M_{\odot}}$ CO WD and a Helium shell of $0.2$ ${\rm M_{\odot}}$
in ref. \cite{wal+11}, which we therefore adopt for our calculations.
Though most of the low mass WD models $({\rm M_{CO-WD}\le0.55}$${\rm M_{\odot}}$)
produce $>2\times10^{-2}$${\rm M_{\odot}}$ $^{44}{\rm Ti}$ (most
produce $>3\times10^{-2}$ ${\rm M_{\odot}}$), consistent with the
overall production rate we use, we do caution that some produce a
few times smaller abundance, which could lead to lower positron production
rates. Models producing as little as $1.5\times10^{-2}\,{\rm M_{\odot}}$
of $^{44}{\rm Ti}$, could still produce the distribution of the 511
keV emission consistent with observations given the larger estimates
for the bulge mass discussed above (up to ${\rm M}_{bulge}=3\times10^{10}\:{\rm M_{\odot}}$),
with even smaller rates still consistent, if the 2005E-like SN rate
are even larger, as expected, given the detection bias discussed earlier
(see section B. above for discussion). Those models producing
significantly less $^{44}{\rm Ti}$ of only a few time $10^{-3}$
${\rm M_{\odot}}$ can explain a significant fraction of the 511 keV
bulge contribution, but can not, by themselves, explain the total
Galactic positron production and the large bulge to disk ratio.

\subsection{Production of positrons in type Ia SNe}

As discussed in depth in refs. \cite{mil+02,pra+11}, standard type
Ia SN suggested to provide a significant source of Galactic positrons
if a non-negligible fraction of the positrons produced in $^{56}{\rm Ni}$
decay escapes the ejecta. However, as already mentioned in the main
text and in refs. \cite{mil+02,pra+11}, the delay time distribution
(and bulge to disk ratio) of regular type Ia SN suggest the vast majority
of these SNe explode in the thin disk and only a very small fraction
of type Ia SNe are expected to currently explode in the old stellar
population of the Galactic bulge. The bulge-concentrated 511 keV line
emission excludes a significant contribution from such sources, and
we therefore do not include such a component in our model.

We should note that in principle fast decaying radioactive elements
such as $^{56}{\rm Ni}$ considered in standard type Ia SNe can also
contribute to positron production in .Ia SNe; again, if significant
fraction of positrons can escape the SN ejecta. Note that the small
amount and lower density of material ejected in such SNe compared
to regular type Ia SNe may give rise to a larger fraction of positron
escape (note that these SNe achieve nebular phase very early, at times
much shorter than the $^{56}{\rm Ni}$ chain decay), and could eventually
contribute to the 511 keV emission. It is therefore possible that
a non-negligible positron production in 05E-like SNe may also arise
from such processes, allowing for smaller positron production from
$^{44}{\rm Ti}$, but these require further study in detailed explosion
models; we do not include any such hypothetical contribution in the
current model discussed here. In that regard it is also interesting
to point out that for .Ia SNe significant abundance of $^{48}{\rm Cr}$
and $^{52}{\rm Fe}$, which are typically unimportant in regular Ia's
can be produced\cite{wal+11}, and can therefore play a similar role
as $^{56}{\rm Ni}$ in providing additional positron sources, if significant
fraction of the positrons can escape the ejecta. 
\end{document}